\documentclass[11pt,a4paper]{article}

\usepackage[T1]{fontenc}
\usepackage[utf8]{inputenc}
\usepackage[russian,english]{babel}
\usepackage[margin=2.6cm]{geometry}
\usepackage{amsmath,amssymb,amsthm}
\usepackage{mathtools}
\usepackage{hyperref}
\hypersetup{colorlinks=true,linkcolor=blue,citecolor=blue,urlcolor=blue}

\newtheorem{theorem}{Theorem}
\newtheorem{lemma}[theorem]{Lemma}

\newtheorem{corollary}[theorem]{Corollary}
\theoremstyle{definition}
\newtheorem{definition}[theorem]{Definition}
\newtheorem{remark}[theorem]{Remark}

\DeclareMathOperator{\DD}{D}
\DeclareMathOperator{\RR}{R}
\newcommand{\N}{\mathbb{N}}
\renewcommand{\Pr}{\mathbf{Pr}}
\newcommand{\free}{\textsf{free}}
\newcommand{\frozen}{\textsf{frozen}}

\title{Enumerating Complexity Revisited}
\author{Alexander Shekhovtsov\thanks{Columbia University, New York, USA.}
\and Nikolay Vereshchagin\thanks{Moscow State University, HSE University,
Yandex. This paper was prepared within the framework of the HSE University Basic
Research Program (project no.~HSE-BR-2025-24).}
\and Georgii Zakharov\thanks{Department of Mathematics and Exeter College,
University of Oxford, Oxford, UK. This research was conducted while the author
was supported by a Hill Foundation Scholarship.}}
\date{}

\begin{document}
\maketitle

\begin{abstract}
We study the relation between deterministic and randomized \emph{enumerating}
Kolmogorov complexity of sets $S\subseteq\N$. Let $\DD(S)$ be the length of a
shortest input on which a fixed optimal deterministic machine enumerates $S$,
and let $\RR(S)$ be the probability that a fixed optimal randomized machine
enumerates $S$. It is easy to see that $-\log\RR(S)\le \DD(S)+O(\log \DD(S))$
for every~$S$; the converse direction is much harder. Solovay proved
$\DD(S)\le -3\log \RR(S)+O(\log\log\frac{1}{\RR(S)})$ for all $S$, while for finite sets
the sharper bound with the constant~$2$ is known.

We show that the case of infinite sets carries no extra cost: \emph{any}
inequality of the form $\DD(F)\le -\alpha\log\RR(F)+O(\log\log\frac{1}{\RR(F)})$ with a
computable~$\alpha$, valid for all finite~$F$, automatically extends to all sets. More generally, we prove a
single dichotomy (Theorem~\ref{th:main}) from which such transfer results follow
for an arbitrary computable non-decreasing bound, and which also shows that every
c.e.\ set  $S$ has a finite subset $F$ such that $F$ has at least as large deterministic complexity as $S$ but randomized one approximately the same or lower. Our result makes  qualitative progress on a problem studied by Solovay in 1977 by showing that the search for the optimal constant can be reduced to a purely combinatorial question about a game played with finite
sets. In particular, our result yields
$\DD(S)\le-2\log\RR(S)+O(\log\log\frac{1}{\RR(S)})$ for every $S$, improving Solovay's
constant $3$ to $2$.   
\end{abstract}

\section{Introduction}

\subsection{Enumerating complexity}

We study Kolmogorov complexity with respect to non-halting Turing machines.
Since such machines do not halt, they can output an infinite object, so instead
of measuring the complexity of strings we measure the complexity of subsets
of~$\N$. There are two natural ways to do this. The \emph{deterministic
enumerating complexity} $\DD(S)$ of a set $S\subseteq \N$ is the length of a
shortest input on which a fixed optimal machine enumerates $S$ (and nothing
else). The \emph{randomized enumerating complexity} is $-\log_2\RR(S)$, where
$\RR(S)$ is the probability that a fixed optimal machine with a random input tape
enumerates exactly~$S$. (All definitions are given in
Section~\ref{sec:prelim}; $\log$ always means $\log_2$.)

Sampling a shortest description immediately gives
\begin{equation}\label{eq:easy}
  -\log \RR(S)\ \le\ \DD(S)+O(\log \DD(S))
\end{equation}
for every $S\subseteq\N$: a random tape begins with a self-delimiting encoding
of a shortest deterministic description of $S$ with probability
$2^{-\DD(S)-O(\log \DD(S))}$ (Lemma~\ref{lem:easy}). The converse direction --- extracting a short
deterministic description from a randomized one --- is the subject of this paper.
It can be viewed as a derandomization problem: given a randomized construction
that outputs the desired object with probability~$p$, we want a deterministic
program of length $O(\log (1/p))$ that reproduces it.

The question goes back to de Leeuw, Moore, Shannon and Shapiro~\cite{dLMSS56},
who proved that $S$ is deterministically enumerable if and only if $\RR(S)>0$;
their argument, however, gives no quantitative bound on $\DD(S)$. Solovay
\cite{Solovay77} proved the linear bound
\begin{equation}\label{eq:solovay}
  \DD(S)\ \le\ -3\log \RR(S)+O(\log\log\frac{1}{\RR(S)})\qquad\text{for all } S\subseteq\N.
\end{equation}
It was later observed that the same method gives the better constant~$2$ for
\emph{finite} sets~\cite{Ver07}. The natural question --- what is the least $\alpha$ for which
$\DD(S)\le-\alpha\log\RR(S)+O(\log\log\frac{1}{\RR(S)})$ holds for all $S$? --- has been
open since then. By~\eqref{eq:easy} we have $\alpha\ge 1$.

In the halting case, the deterministic complexity equals the randomized complexity up to additive logarithmic terms due to the well-known coding theorems (\cite{Levin74}). These theorems significantly use the fact that a machine eventually halts. If a machine has halted, then the output it has produced is fixed. In contrast, in the non-halting setting that we consider, the output can change indefinitely as there is no fixed finishing time. This difference makes the problem for the non-halting case harder. 

\subsection{Our results}

Our main result is a single statement about the interplay of the two
complexities, which we state first because everything else follows from it. For
$k,l\in\N$ we write $\RR(F\mid k,l)$ for the conditional a priori probability of
enumerating $F$ given the pair $(k,l)$ (Definition~\ref{def:condrand}).

\begin{theorem}[Main dichotomy]\label{th:main}
There is a constant $d$ such that for all $k,l\in\N$ and every $S\subseteq \N$
with $\RR(S)\ge 2^{-k}$ at least one of the following holds:
\begin{enumerate}
\item[(i)] $\DD(S)\le l$;
\item[(ii)] there is a \emph{finite} $F\subseteq S$ with $\DD(F)>l$ and
  $\RR(F\mid k,l)\ge 2^{-k}/d$.
\end{enumerate}
\end{theorem}

Equivalently: if every finite $F\subseteq S$ with $\RR(F\mid k,l)\ge 2^{-k}/d$
satisfies $\DD(F)\le l$, then $\RR(S)\ge2^{-k}$ implies $\DD(S)\le l$. In other
words,
\emph{the only obstruction to a short deterministic description of $S$ is a
finite subset of $S$ that is itself hard to describe deterministically but easy
to hit randomly}. It is essential that the two parameters $k$ and $l$ appear as a
condition in~(ii): the machine we construct receives them as input, so the
constant $d$ does not depend on them. This is what makes the theorem
loss-free, and it is the point where our argument departs from
Solovay's~\cite{Solovay77}.

The transfer of upper bounds from finite to arbitrary sets follows at once.

\begin{theorem}[Transfer]\label{th:transfer}
Let $f\colon\N\to\N$ be a non-decreasing computable function such that
\[
  \DD(F)<f(k)\quad\text{for every $k$ and every \emph{finite} $F\subseteq\N$ with }
  \RR(F)\ge2^{-k}.
\]
Then there is a constant $C$ such that
\[
  \DD(S)<f\bigl(k+2\log k+C\bigr)\quad\text{for every $k$ and every 
\emph{infinite} }
  S\subseteq \N \text{ with } \RR(S)\ge 2^{-k}.
\]
\end{theorem}

Specializing $f$ to a linear function we obtain a simple corollary on the transfer
from finite to infinite sets, with no loss in the leading constant.

\begin{corollary}\label{cor:alpha}
Let $\alpha\ge1$ be a computable real such that
$\DD(F)\le-\alpha\log\RR(F)+O(\log\log\frac{1}{\RR(F)})$ for every \emph{finite} $F\subseteq\N$.
Then $\DD(S)\le-\alpha\log \RR(S)+O(\log\log\frac{1}{\RR(S)})$ for every \emph{infinite} $S\subseteq\N$
(with a different constant in the $O$-term).
\end{corollary}

Together with the bound for finite sets from~\cite{Ver07} this improves
Solovay's constant~$3$ in~\eqref{eq:solovay} to~$2$.

\begin{corollary}\label{cor:two}
For every $S\subseteq\N$ we have
$\DD(S)\le-2\log \RR(S)+O(\log\log\frac{1}{\RR(S)})$.
\end{corollary}

Finally, Theorem~\ref{th:main} applied to the parameters of $S$ itself says that
every c.e.\ set is faithfully represented by one of its finite subsets.

\begin{corollary}\label{cor:hardfinite}
For every $S\subseteq\N$ with $\RR(S)>0$ there is a finite $F\subseteq S$ with
\[
  \DD(F)\ \ge\ \DD(S)
  \qquad\text{and}\qquad
  -\log\RR(F)\ \le\ -\log\RR(S)+O\Bigl(\log\log\tfrac{1}{\RR(S)}\Bigr).
\]
\end{corollary}

In other words, the deterministic hardness of an arbitrary set is always
witnessed by one of its finite subsets, and the randomized machine still hits that
subset with essentially the probability it had for $S$. Theorem~\ref{th:main}
gives this first in conditional form, namely $\RR(F\mid k,l)\ge\RR(S)/d$ for
$k=\lceil\log(1/\RR(S))\rceil$ and $l=\DD(S)-1$; here the condition can then be
removed for free, because $l\le 2k+O(\log k)$ by Corollary~\ref{cor:two}, so that
specifying the pair $(k,l)$ costs only $O(\log k)$ bits. 

\subsection{Comparison with previous work}

Solovay~\cite{Solovay77} also proceeds by reducing infinite sets to finite ones,
but  Solovay's reduction loses an additive~$1$ in the constant.  In order to approximate $\RR(A)$ for an infinite $A$ one truncates the universe to $[n] = \{0, 1, \ldots, n\}$ and also terminates the machine after $t$ steps. More precisely, define $\RR^t_n(A)$ as the probability that the optimal randomized machine enumerates a set $S$, such that $S \cap [n] = A\cap[n]$, if stopped after $t$ steps. In the case $t = \infty$, we consider the set $S$ that the machine enumerates in the limit. It can be shown that $\RR_n^t(A) \to \RR_n^{\infty}(A)$ as $t$ goes to infinity, and also that $\RR_n^{\infty}(A) \to \RR(A)$ as $n$ goes to infinity. Using these properties, we can try to find $t_n$, such that $\RR_n^{t_n}(A)$ is a good approximation of $\RR_n^{\infty}(A)$, and then we would have $\RR_n^{t_n}(A) \to \RR(A)$ as $n$ goes to infinity, allowing us to estimate $\RR(A)$. However, specifying the sequence $(t_n)$
costs an extra $-\log \RR(A)$ bits. Therefore, Solovay's result requires an extra $-\log \RR(S)$ term:

\begin{theorem}[\cite{Solovay77}]\label{th:solovay-red}
Let $\alpha\ge1$ be such that $\DD(F)\le-\alpha\log\RR(F)+O(\log\log\frac{1}{\RR(F)})$ for
all finite $F\subseteq\N$. Then
$\DD(S)\le-(\alpha+1)\log\RR(S)+O(\log\log\frac{1}{\RR(S)})$ for all $S\subseteq\N$.
\end{theorem}

Corollary~\ref{cor:alpha} removes the additive~$1$. Our proof avoids using the sequence $(t_n)$. Instead, we develop a different technique. For a parameter $k \in \mathbb{N}$, we search for sets $A$ such that $\RR(A) \geq 2^{-k}$. We design another randomized machine $N$ with the property that for any finite $A' \subseteq A$, there will be a finite set $C$, such that $A' \subseteq C \subseteq A$ and $N$ enumerates $C$ with probability at least $2^{-k - O(1)}$ at some step of its execution. Essentially, $N$ witnesses that the initial problem can be reduced to the finite sets case.

The problem of finding the optimal $\alpha$ for finite sets is known to be
equivalent to a purely combinatorial question about a game. Solovay's method for
finite sets rests on a winning strategy in \emph{Martin's game}, which has one
parameter $\varepsilon\in(0,1)$; writing $N(\varepsilon)$ for the least number of
sets needed to win it, Solovay proved $N(\varepsilon)=O((1/\varepsilon)^2)$ and
Ageev~\cite{Ageev02} proved the matching lower bound
$N(\varepsilon)=\Omega((1/\varepsilon)^2)$. Thus the constant~$2$ cannot be
improved along this route. A related game, \emph{Cats versus Ants}, was
introduced in~\cite{Ver08}: let $K(\varepsilon)$ be the least number of cats
needed to win it. Every strategy for Martin's game yields one for Cats versus
Ants, so $K(\varepsilon)\le N(\varepsilon)$, but Ageev's lower bound does not
carry over; only the trivial bound $K(\varepsilon)=\Omega(1/\varepsilon)$ is
known. The relevance of the game is that it controls the optimal $\alpha$ for
finite sets from both sides: a computable strategy for Alice with
$c(1/\varepsilon)^{\alpha}$ cats bounds $\alpha$ from above, and computable
winning strategies for Bob bound it from below (Theorem~\ref{th:game} below).
By Corollary~\ref{cor:alpha}  any computable bound on finite sets holds
for arbitrary ones, so the question ``how much do random bits help in
non-halting Kolmogorov complexity?'' becomes a combinatorial problem about cats
and ants, with the answer known to lie between $1$ and~$2$. The game
is defined, and the reduction stated precisely, in Section~\ref{sec:game}.

\subsection{The proof idea}\label{sec:idea}

The proof of Theorem~\ref{th:main} is
naturally described in terms of cats and ants, and we will use this language
throughout Section~\ref{sec:proof}; the reader who prefers a formal description
will find one there as well.

We think of the finite subsets of $\N$ as the vertices of a graph, where an
enumeration process is a token moving \emph{upward} (a set only grows). We simulate the
optimal randomized  machine $P$ on all  outcomes of the random tape simultaneously.
Let us think of the outcome $w \in \{0, 1\}^{\mathbb{N}}$ of the random tape as an \emph{ant} of infinitesimal
mass, sitting at the finite set enumerated by $P$ on $w$ so far. The total mass of
ants whose limit position is $S$ is $\RR(S)$. We run, in parallel, the optimal
deterministic machine $M$ on all inputs of length at most~$l$. Let us think of each such
input as a \emph{cat}, sitting at the finite set it has enumerated so far. There
are fewer than $2^{l+1}$ cats. Our goal is to show that if the good ants (those
converging to $S$) have mass at least $2^{-k}$, then either some cat converges to
$S$ --- giving $\DD(S)\le l$ --- or some finite set $F \subseteq S$ has $2^{-k - O(1)}$ mass of ants
and no cat in the limit.

To conduct our argument, we define a randomized machine $N$. It is defined in stages $t=1, 2, \ldots$. At stage $t$ we run the first $t$
steps of $P$ on every outcome of the random tape and the first $t$ steps of $M$ on every
input of length at most $l$; the \emph{position} of an ant or of a cat at stage
$t$ is the finite set enumerated so far by the corresponding run. Positions 
therefore only grow with $t$.  Let us construct $N$ by referring to outcomes of its random tape as \textit{slow ants}.  In other words, each ant $w$ that corresponds to $P$ gets a
\emph{slow copy} that corresponds to $N$. The slow copy always enumerates a subset of what the original ant enumerates. Each slow copy is in one of two states at every
moment: either \emph{free} or \emph{frozen}.  Note that the behaviours of the slow copies exactly define the behaviour of $N$ and vice versa. The machine $N$ also receives $k$ and $l$ as inputs. Write $\RR'(S)=\RR_N(S)$ for the probability that $N$ enumerates a given set $S$. For each stage $t$, there are two rules that govern the movement of the slow copies.

\emph{Collecting.} If some finite set $F$ has the property that free slow ants of
total mass at least $2^{-k-2}$ sit below $F$ (enumerate a subset of $F$) while their fast copies have
already passed above $F$ (enumerate a superset of $F$), then all these slow ants are moved to $F$ and frozen
there.  This action allows us to concentrate slow ants of total mass at least $2^{-k-2}$  in the set $F$.

\emph{Scattering.} If a cat sits exactly at a set $F$ where frozen slow ants
sit, they are released: each of them jumps to the current position of its fast
 copy and becomes free again.

Now look at a good ant. Its slow copy can be scattered infinitely often, or it
can be frozen from some moment on, or it can end up free forever. For at least one good ant, 
the last possibility is excluded by a continuity-of-measure argument: if all good slow ants
 became free
forever, then from some moment on for some natural $n$, a mass of at least $2^{-k-2}$ of them would
sit below $S \cap [n]$  with their fast  copies above
$S \cap [n]$, and the collecting rule would freeze them. 
So pick a good ant that is scattered infinitely often or frozen from some moment on.
In the first case, a single cat
scatters our slow ant infinitely often; at each scattering the cat's position
coincides with the slow ant's position, which is squeezed between the previous
position of the fast ant and $S$; hence that cat enumerates exactly $S$ and
$\DD(S)\le l$. In the second case, the slow ant is frozen forever at some finite
$F\subseteq S$; no cat ever sits at $F$ afterwards, so $\DD(F)>l$; and the whole
batch that was frozen together with it --- of mass at least $2^{-k-2}$ --- stays
there forever, so $\RR'(F\mid k,l)\ge2^{-k-2}$, whence $\RR(F\mid k,l)\ge
2^{-k}/d$ by the optimality of $\RR$. 

\section{Preliminaries}\label{sec:prelim}

Throughout, $\log$ is the binary logarithm. Further, $\langle\cdot\rangle$ denotes a fixed self-delimiting encoding
with $|\langle y\rangle|\le|y|+2\log|y|+O(1)$ for strings and $|\langle
k\rangle|\le 2\log k+O(1)$ for natural numbers.  Write $[n] = \{0, 1, 2, \ldots, n \}$. We identify a natural number with its binary representation.

\begin{definition}[Deterministic enumerating complexity]\label{def:det}
Let $M$ be a Turing machine with one (read only) input tape and one (write only) output tape which  possibly does not halt. The input tape contains a finite string $p\in\{0,1\}^*$ padded with a
symbol outside $\{0,1\}$, so the machine can detect where $p$ ends. We say that
$M$ \emph{enumerates} $S\subseteq\N$ on the input $p$ if the set of numbers ever
written on the output tape is exactly $S$. Put
$\DD_M(S)=\min\{|p|: M \text{ enumerates } S \text{ on } p\}$
($=\infty$ if there is no such $p$). There is an \emph{optimal} machine $M^{*}$, i.e.\
one with $\DD_{M^{*}}(S)\le \DD_M(S)+O(1)$ for every $M$ and all $S$; we fix such
an $M^{*}$ and put $\DD:=\DD_{M^{*}}$.   
\end{definition}

\begin{definition}[Conditional deterministic complexity]\label{def:detcond}
Let $M$ have two input tapes; a string $p$ is written on the first one and a
string $y$ on the second, both padded as in Definition~\ref{def:det}.
Put $\DD_M(S\mid y)=\min\{|p|:M \text{ enumerates } S \text{ on } (p,y)\}$. Again
there is an optimal machine, and $\DD(S\mid y)$ is defined up to an additive
constant.
\end{definition}

\begin{definition}[Randomized enumerating complexity]\label{def:rand}
Let $P$ be a machine with one infinite input tape and one output tape. For
$w\in\{0,1\}^{\N}$ let $P(w)$ be the set enumerated by $P$ when $w$ is written on
the input tape.
With respect to the uniform measure on $\{0,1\}^{\N}$ put
$\RR_P(S)=\Pr_{w}[P(w)=S]$.\footnote{The event $\{w:P(w)=S\}$ is measurable: for
each $n$ the event ``$n$ is eventually enumerated'' is a countable union of
cylinders, and $\{w:P(w)=S\}$ is the intersection of these events for $n\in S$
and of their complements for $n\notin S$. The same applies to all sets of tapes
considered in Section~\ref{sec:proof}, which are obtained from events depending
on finitely many bits by countable unions and intersections.} There is an optimal machine $P^{*}$, i.e.\ one with
$\RR_{P^{*}}(S)\ge c_P\cdot\RR_P(S)$ for every $P$ and all $S$, with $c_P>0$
depending only on $P$; we fix such a $P^{*}$ and put $\RR:=\RR_{P^{*}}$. The
randomized enumerating complexity of $S$ is $-\log \RR(S)$.
\end{definition}

\begin{definition}[Conditional randomized complexity]\label{def:condrand}
Let $P$ have an infinite random input tape and a second input tape carrying a string
$y$ (padded as in Definition~\ref{def:det}), and put $\RR_P(S\mid y)=\Pr_{w}[P(w,y)=S]$. Again an optimal machine exists
and $\RR(S\mid y)$ is defined up to a multiplicative constant. For natural
numbers $k,l$ we write $\RR(S\mid k,l):=\RR(S\mid \langle k\rangle l)$.
\end{definition}

The existence of optimal machines for all four definitions can be proven by a standard argument,
see e.g.~\cite{LV08}.

 Without loss of generality, let us assume that $\DD(S) > 0$ and $\RR(S) \leq \frac14$ for any $S$. Indeed, since the events
$\{w:P^{*}(w)=S\}$ are disjoint, $\sum_S\RR(S)\le1$, so at most  four sets have
$\RR(S)>1/4$; and only the empty input can witness $\DD(S)=0$, so at most one
set has $\DD(S)=0$. All the bounds below are of the form $\DD(S)\le\dots+O(1)$,
so for these  at most five particular sets they hold after enlarging the additive
constant.  Thus, whenever we write $\log\DD(S)$ or
$\log\log\frac1{\RR(S)}$, we have $\DD(S)\ge1$ and $\RR(S)\le1/4$, so that both
expressions are defined and non-negative.
All optimal machines above are fixed once and for all, and the constants in all terms $O(m)$ below  depend only on the choice of optimal machines.

\begin{remark}[Notation used elsewhere]\label{rem:notation}
Solovay~\cite{Solovay77} and~\cite{Ver08}, following Chaitin~\cite{Chaitin75},
write $I(A)$ for
the complexity of the problem ``enumerate $A$'' (our $\DD(A)$), $m(A)$ for the a
priori probability of enumerating $A$ (our $\RR(A)$), and
$H(A)=\lceil-\log m(A)\rceil$ (our $-\log \RR(A)$). 
One difference should be kept in mind: the complexity of~\cite{Solovay77,Ver08}
is a \emph{prefix} one --- there $p$ is required to enumerate $A$ on every
infinite extension of $p$ --- whereas our $\DD$ is a \emph{plain} one, since in
Definition~\ref{def:det} the machine is told where its input ends. The two are
close: $\DD(A)\le I(A)+O(1)$ (a machine that is told where its input ends can
simulate a prefix machine on it), and $I(A)\le\DD(A)+2\log\DD(A)+O(1)$ (encode
the input self-delimitingly), exactly as for the plain and the prefix complexity
of strings~\cite{LV08,SUV}. Since $\DD(A)=O\bigl(\log(1/\RR(A))\bigr)$ by
Corollary~\ref{cor:two}, the two conventions differ by
$O\bigl(\log\log(1/\RR(A))\bigr)$, so every statement in this paper is
insensitive to the choice.
\end{remark}

The following two facts are standard.

\begin{lemma}\label{lem:cond-basic}
For all $S\subseteq \N$, all strings $y$ and all $k\in\N$:
\begin{enumerate}
\item[(a)] $\RR(S\mid y)\ \ge\ c\cdot\RR(S)$;
\item[(b)] $\RR(S)\ \ge\ c\cdot 2^{-|\langle y\rangle|}\RR(S\mid y)$; in
  particular $\RR(S)\ge c\,k^{-2}\RR(S\mid k)$;
\item[(c)] if a string $y$ is computed from a string $x$ by an algorithm $g$,
  then
  \[
    \RR(S\mid x)\ \ge\ 2^{-m-O(1)}\cdot \RR(S\mid y)
    \qquad\text{and}\qquad
    \DD(S\mid x)\ \le\ \DD(S\mid y)+m+O(1),
  \]
  where $m$ is the length of a program computing $g$ and the constants in
  $O(1)$ do not depend on $g$.
\end{enumerate}
Here $c>0$ is an absolute constant.
\end{lemma}

These are the obvious analogues, for enumerating complexity, of well-known
properties of Kolmogorov complexity and of the a priori probability, and they are
proved in exactly the same way; see~\cite{LV08,SUV}.

\begin{lemma}\label{lem:easy}
$-\log \RR(S)\le \DD(S)+2\log\DD(S)+O(1)$ for every $S\subseteq\N$.
\end{lemma}

\begin{proof}
Let $p$ be a shortest input on which the machine $M^{*}$ of
Definition~\ref{def:det} enumerates $S$, so that $|p|=\DD(S)$. Consider the
randomized machine $P$ that reads from its random tape a self-delimiting code
$\langle p\rangle$ and then simulates $M^{*}$ on the input $p$ (having read
$\langle p\rangle$, it knows $p$ completely and in particular knows where $p$
ends). The tapes beginning with $\langle p\rangle$ have measure
$2^{-|\langle p\rangle|}\ge2^{-|p|-2\log|p|-O(1)}$, and $P$ enumerates $S$ on
every such tape, so $\RR_P(S)\ge2^{-\DD(S)-2\log\DD(S)-O(1)}$.

It remains to pass from $P$ to the optimal randomized machine $P^{*}$ of
Definition~\ref{def:rand}, which has nothing to do with $M^{*}$: by the optimality
of $P^{*}$ we have $\RR(S)=\RR_{P^{*}}(S)\ge c_P\RR_P(S)$ for a
constant $c_P>0$ depending only on $P$, which is the assertion of the lemma.
\end{proof}

\section{Proof of the main dichotomy}\label{sec:proof}

Fix $k,l\in\N$ and let $S\subseteq \N$ satisfy $\RR(S)\ge2^{-k}$. If $S$ is
finite the statement is trivial: either $\DD(S)\le l$, or $F:=S$ witnesses~(ii),
because $\RR(S\mid k,l)\ge c\RR(S)\ge c2^{-k}$ by
Lemma~\ref{lem:cond-basic}(a), and it suffices to take $d\ge 1/c$. So assume
from now on that $S$ is infinite.

\subsection{The process}\label{sec:process}

We describe a randomized machine $N$, as in Definition~\ref{def:condrand},
which receives $\langle k\rangle l$ on the second input tape. 
On its output tape the machine writes its \emph{output so far}. On its
work tape it maintains, among other things,  bits
$\sigma(w)\in\{\free,\frozen\}$. 

Besides, the machine $N$ simulates on its work tape 
 the optimal randomized machine $P^{*}$ from Definition~\ref{def:rand}
and  the optimal deterministic machine $M^{*}$ from Definition~\ref{def:det}
on all inputs $p$ of length at most $l$ in a dovetail fashion. 
This means the following.

Both machines $P^*,M^*$ work in steps. Let $P^{*\,t}(w)$ be the finite set enumerated by $P^*$ on random input $w$ within $t$ steps.
The machine reads at most one bit of its random tape per step, so $P^{*\,t}(w)$
depends only on $w_1\dots w_t$ , and, therefore, can be simulated. 
Likewise for a string $p$ of length at most $l$ let $M^{*\,t}(p)$ denote
the set of natural $n$ printed by $M^{*}$ on the input $p$ within $t$ steps.
The set $M^{*\,t}(p)$ does not decrease as $t$ increases and its union over all
$t$ is $M^{*}(p)$, the set enumerated by $M^{*}$ on $p$.

The machine $N$ works in stages $t=0,1,\dots$.
On stage $t$ it simulates $t$ steps of   $P^{*}$ on all $w$
and $t$ steps of  $M^{*}$ on all inputs $p$ of length at most $l$.
As already pointed out, the action of $P^*$ within $t$ steps
depends only on the length-$t$ prefix of $w$, so these simulations can be performed in finite time.
We call  $w$, viewed as a point of the probability
space, a \emph{fast ant} sitting at $P^{*\,t}(w)$ and
call $p$ a \emph{cat} sitting at $M^{*\,t}(p)$. There are fewer than $2^{l+1}$ cats.

In a similar way we will denote by $N^t(w)$ the output of $N$ by the end of
stage $t$ on random input $w$. This set $N^t(w)$ is for us to define. It must be finite and must not decrease as $t$ increases.

The pair $(N^{t}(w),\sigma_t(w))$ will be viewed as the position and state of the
\emph{slow ant} $w$. Initially $N^{0}(w)=\varnothing$ and $\sigma_0(w)=\free$ for
all $w$. At each stage $t=1,2,\dots$ the machine $N$ advances the simulations of
$P^{*},M^*$ by one step, and then performs the following two
operations in this order. Neither operation depends on the random input of $N$.

\medskip
\noindent At the beginning of stage $t$ set $B(w):=N^{t-1}(w)$ and
$\sigma(w):=\sigma_{t-1}(w)$ for all $w$. These are the current position and
state of the slow ant $w$; both operations below change them, and their values at
the end of the stage will be taken as $N^{t}(w)$ and $\sigma_t(w)$.

\medskip
\noindent\textbf{Scattering.} For every finite set $F$ such that
$M^{*\,t}(p)=F$ for some cat $p$: for every $w$ such that $\sigma(w)=\frozen$ and
$N^{t-1}(w)=F$, set $B(w):=P^{*\,t}(w)$ and $\sigma(w):=\free$.
(If a cat sits exactly at a set $F$ where the frozen slow ant $w$
sits, the ant $w$ is released: it jumps to the current position of its fast
copy and becomes free again.) 

\medskip
\noindent\textbf{Collecting.} 
Do the following loop.
While there exists a finite set $F$ with
\begin{equation}\label{eq:collect}
  \Pr\bigl[\,w:\ \sigma(w)=\free,\ B(w)\subseteq F\subseteq P^{*\,t}(w)\,\bigr]\ \ge\ 2^{-k-2},
\end{equation}
pick such an $F$ and, for every $w$ counted in~\eqref{eq:collect}, set
$B(w):=F$ and $\sigma(w):=\frozen$. 
It might happen that there are several such $F$; it does not matter which one is picked.
(If some finite set $F$ has the property that free slow ants of
total mass at least $2^{-k-2}$ sit below $F$ while their fast  copies have
already passed above $F$, then all these slow ants are moved to $F$ and frozen
there.) 

Note that this loop can be iterated only finitely many times.
Indeed, each iteration turns a set of free slow ants of mass at least $2^{-k-2}$ into
frozen ones, and no ant is unfrozen during the loop. As the total mass is $1$,
there are at most $2^{k+2}$ iterations.
Also note that neither operation depends on $N$'s own random input: the work of
$N$ is almost independent of it, and the only thing that does depend on it is the
output of $N$.

Once both operations have been done, we set $\sigma_t(w):=\sigma(w)$; then the
machine $N$ reads the $t$-th bit of its own random input $w$
and  prints on its output tape 
all numbers from $B(w)\setminus N^{t-1}(w)$. Obviously we have $N^{t-1}(w)\subseteq B(w)$,
so the set enumerated by the end of the stage is $N^{t}(w)=B(w)$.
Note that both $B(w)$ and $\sigma(w)$
depend only on length-$t$ prefix of $w$, thus there is no need to read more random bits.

\medskip
Thus $N$ enumerates, on the random tape $w$, the set
$N(w)=\bigcup_t N^{t}(w)$. We record the elementary properties of the process.

\begin{lemma}\label{lem:process}
\begin{enumerate}
\item[(a)] $N^{t}(w)\subseteq P^{*\,t}(w)$ for all $t,w$, and $N^{t}(w)$ is
  non-decreasing in $t$; hence $N$ is a legitimate enumerator and
  $N(w)\subseteq P^{*}(w)$.
\item[(b)] For every $t$, the functions $w\mapsto (P^{*\,t}(w),N^{t}(w),\sigma_t(w))$
  depend only on $w_1\dots w_t$. In particular all the probabilities
  in~\eqref{eq:collect} are computable rationals with denominator $2^{t}$, and
  the process is computable given $k$ and~$l$.
\end{enumerate}
\end{lemma}

\begin{proof}
(a) At a stage $t$ the position of a slow ant either stays the same or is changed by
one of the two operations, and we check the claim in all three cases. If the ant
is not moved at all, its position remains $N^{t-1}(w)\subseteq
P^{*\,(t-1)}(w)\subseteq P^{*\,t}(w)$ by the induction hypothesis and the
monotonicity of $P^{*\,t}(w)$ in $t$. Scattering puts the ant at $P^{*\,t}(w)$
itself, and collecting puts it at a set $F$ with $B(w)\subseteq F\subseteq
P^{*\,t}(w)$. So in each case the new position contains the old one and is
contained in $P^{*\,t}(w)$, and the operations of a single stage can only be
applied one after another, each of them preserving both properties.

(b) Induction on $t$. The set $P^{*\,t}(w)$ depends only on the first $t$ bits of
$w$ because $P^{*}$ reads at most one bit per step. The conditions in the two
operations are stated in terms of $P^{*\,t}$, $B$, $\sigma$, $N^{t-1}$ and the
(deterministic) positions of the cats, and both operations act identically on all
$w$ satisfying the same condition. Hence the property is preserved. Consequently
every set occurring in~\eqref{eq:collect} is a finite union of cylinders of rank
$t$. Note also that $F$ has to satisfy $F\subseteq P^{*\,t}(w)$ for at least one
$w$, so it is a subset of the finite set $\bigcup_{|u|=t}P^{*\,t}(u)$, and the
loop searches through finitely many candidates.
%
\end{proof}

\begin{remark}
Formally, a machine cannot store the uncountable objects $N(\cdot)$,
$\sigma(\cdot)$. By Lemma~\ref{lem:process}(b) it does not have to: at stage $t$
it keeps the finite table of the values of $(B,\sigma)$ on the $2^{t}$ cylinders
of rank $t$, and when passing from $t$ to $t+1$ it copies the value at a cylinder
$u$ to the two cylinders $u0$ and $u1$. On the random input $w$ actually supplied to $N$,
the machine prints the elements of $N(w)$ as they appear.
\end{remark}

Call an ant $w$ (fast or slow) \emph{good} if $P^{*}(w)=S$. By assumption the
good fast ants have mass $\RR(S)\ge 2^{-k}$. Note that a slow ant is scattered only if
it is frozen, and it is frozen only by collecting, which requires it to be free;
so the states alternate,
\[
  \free\ \xrightleftharpoons[\ \text{scattering}\ ]{\ \text{collecting}\ }\ \frozen
\]
(collecting freezes a free ant, scattering frees a frozen one). Within a single
stage $t$ an ant can change its state twice, from $\frozen$ to $\free$ by
scattering and back to $\frozen$ by collecting, but no more: the two operations
are performed once each and in this order. In particular, an ant with
$\sigma_{t-1}(w)=\free$ is not touched by scattering, so during stage $t$ its
state either does not change at all or changes once, from $\free$ to $\frozen$;
we use this in the proof of Lemma~\ref{lem:notC}.

Each slow ant satisfies exactly one of:
\begin{enumerate}
\item[(A)] it is scattered (and collected) infinitely often;
\item[(B)] from some moment on it is frozen forever;
\item[(C)] from some moment on it is free forever.
\end{enumerate}
Indeed, if $w$ is scattered finitely often then after the last scattering it is
free, and it can be frozen at most once more (a second freezing would require an
intervening scattering); so it stays free forever or is frozen forever. In case
(C) the slow ant never moves after the last scattering, because both operations
that move it change its state.

\subsection{Not all good ants are eventually free}

\begin{lemma}\label{lem:notC}
Not all good slow ants are in case~(C).
\end{lemma}

\begin{proof}
Assume they all are. Then each good slow ant $w$ has a limit position $N(w)$, a
finite subset of $S$ by Lemma~\ref{lem:process}(a), which it reaches at some
finite moment and never leaves, while being free from that moment on.

For $n\in\N$ let $G_n:=\{w \text{ good}: N(w)\subseteq S \cap [n]\}$ .
As shown above, the limit position $N(w)$ of every good slow ant is a finite
subset of $S$ and hence is contained in $S \cap [n]$ for some $n$; that is,
the union of the sequence $(G_n)$  is equal to the
set of all good ants, of mass at least $2^{-k}$.
Since
the sequence $(G_n)$ is non-decreasing, by continuity of measure there is an $n$ with
\[
  \Pr[G_n]\ \ge\ 2^{-k-1}.
\]
Fix this $n$ and put $F:=S \cap [n]$. For $t\in\N$ let
\[
  E_t:=\{w\in G_n:\ \text{for all } t'\ge t,\
  P^{*\,t'}(w)\supseteq F,\ \sigma_{t'}(w)=\free \text{ and } N^{t'}(w)=N(w)\}.
\]
For every $w\in G_n$ we have $P^{*\,t'}(w)\uparrow S\supseteq F$ and, by our
assumption, $w$ is eventually free; hence $w\in E_t$ for all large
$t$. Thus $(E_t)$ is non-decreasing with union $G_n$, and by continuity of
measure there is a $t_0$ with
\[
  \Pr[E_{t_0}]\ \ge\ 2^{-k-2}.
\]
Consider the collecting loop at any stage $t\ge t_0$ and look at its final state,
when no $F'$ satisfies~\eqref{eq:collect}. By the definition of $E_{t_0}$, every
$w\in E_{t_0}$ is free at the end of stage $t$ and sits at $N(w)$. The loop only
freezes ants and never frees them, so an ant free at the end of the loop was free
throughout it and was never moved by it; note that this argument runs backwards
from the end of the stage, so it applies to $t=t_0$ as well, although
$\sigma_{t_0}(w)=\free$ is only known at the end of stage~$t_0$. Thus at the end of the loop every $w\in
E_{t_0}$ is free and its current position satisfies $B(w)=N(w)\subseteq
S\cap[n]=F\subseteq P^{*\,t}(w)$. Hence
$F=S\cap[n]$ satisfies~\eqref{eq:collect} with mass at least $\Pr[E_{t_0}]\ge2^{-k-2}$, so
the loop could not have stopped --- a contradiction.
\end{proof}

\subsection{A good ant scattered infinitely often}

\begin{lemma}\label{lem:caseA}
If some good slow ant is in case~(A), then $\DD(S)\le l$.
\end{lemma}

\begin{proof}
Let $w_0$ be such an ant and let $t_1<t_2<\cdots$ be the stages at which it is
scattered. The scattering at stage $t$ requires a cat sitting at the position the
slow ant had at the end of stage $t-1$; as there are fewer than $2^{l+1}$ cats,
by the pigeonhole principle there is a single cat $p$ (thus $|p|\le l$) and an
infinite subsequence, which we again denote $t_1<t_2<\cdots$, such that
\[
  M^{*\,t_i}(p)\ =\ N^{t_i-1}(w_0)\ =:\ F_i \qquad (i=1,2,\dots).
\]
Immediately after the scattering at stage $t_i$ the slow ant sits at
$P^{*\,t_i}(w_0)$, and its position never decreases, so
\[
  F_{i+1}\ =\ N^{t_{i+1}-1}(w_0)\ \supseteq\ P^{*\,t_i}(w_0).
\]
Since $w_0$ is good, $P^{*\,t}(w_0)\uparrow S$, and therefore $\bigcup_i
F_i\supseteq\bigcup_i P^{*\,t_i}(w_0)=S$. On the other hand
$F_i=N^{t_i-1}(w_0)\subseteq P^{*\,t_i-1}(w_0)\subseteq S$ by
Lemma~\ref{lem:process}(a). Hence
\[
  \bigcup_i F_i\ =\ S .
\]
It remains to note that the cat $p$ enumerates exactly $S$. Indeed,
$M^{*}(p)=\bigcup_t M^{*\,t}(p)\supseteq\bigcup_i F_i=S$; conversely, for every
$t$ there is $t_i>t$, and by monotonicity $M^{*\,t}(p)\subseteq
M^{*\,t_i}(p)=F_i\subseteq S$, so $M^{*}(p)\subseteq S$. Thus $M^{*}(p)=S$ and
$\DD(S)\le|p|\le l$.
\end{proof}

\subsection{A good ant frozen forever}

\begin{lemma}\label{lem:caseB}
If some good slow ant is in case~(B), then there is a finite $F\subseteq S$ with
$\DD(F)>l$ and $\RR_N(F\mid k,l)\ge 2^{-k-2}$.
\end{lemma}

\begin{proof}
Let $w_0$ be such an ant, let $t_0$ be the stage of its last freezing and let
$F:=N^{t_0}(w_0)$ be its position from $t_0$ on. Then $F$ is a finite subset of
$S$ by Lemma~\ref{lem:process}(a).

First, $\DD(F)>l$. 
Indeed, assume that $M^{*}(p)=F$ for some $p$ with
$|p|\le l$; then $p$ is one of the cats. As $F$ is finite and $M^{*\,t}(p)$ is
non-decreasing with union $F$, we have $M^{*\,t}(p)=F$ for all sufficiently
large $t$ --- so the cat $p$ sits at $F$ at some stage $t\ge t_0+1$.
The scattering rule
at stage $t$ would release $w_0$, which sits at $F$ at the end of stage $t-1$, a
contradiction. 

Second, consider the collecting operation which froze $w_0$ at stage $t_0$. It
froze at $F$ the whole set
$W:=\{w:\sigma(w)=\free,\ B(w)\subseteq F\subseteq P^{*\,t_0}(w)\}$, where
$B$ and $\sigma$ are the current values immediately before that operation; the
mass of $W$ is at least $2^{-k-2}$ by~\eqref{eq:collect}. Every $w\in W$ sits at
$F$ in the frozen state after stage $t_0$, and can be released only by a cat
sitting at $F$. That never happens: a cat sitting at $F$ at a stage $t\ge t_0+1$
would, by the scattering rule, release $w_0$ as well, contradicting the choice of
$w_0$. (Note that this does not follow from $\DD(F)>l$ alone: a cat may visit $F$
in passing, with $M^{*\,t}(p)=F$ at some stage and $M^{*}(p)\supsetneq F$ in the
limit.) Hence $N^{t}(w)=F$ for all $t\ge t_0$ and all
$w\in W$, so $N(w)=F$ and $N$ enumerates $F$ on every $w\in W$:
\[
  \RR_N(F\mid k,l)\ \ge\ \Pr[W]\ \ge\ 2^{-k-2}. \qedhere
\]
\end{proof}

\subsection{Putting the pieces together}

\begin{proof}[Proof of Theorem~\ref{th:main}]
The case of finite $S$ was settled at the beginning of the section, so let $S$ be
infinite with $\RR(S)\ge 2^{-k}$ and run the process of Section~\ref{sec:process}
with the parameters $k,l$. By Lemma~\ref{lem:notC} some good slow ant is in case
(A) or in case~(B).

If some good slow ant is in case~(A), then $\DD(S)\le l$ by
Lemma~\ref{lem:caseA} and alternative~(i) holds.

Otherwise some good slow ant is in case~(B), and Lemma~\ref{lem:caseB} provides a
finite $F\subseteq S$ with $\DD(F)>l$ and $\RR_N(F\mid k,l)\ge2^{-k-2}$. The
machine $N$ is one fixed machine which receives $k$ and $l$ on its second input
tape; hence by optimality of $\RR(\cdot\mid\cdot)$ there is an absolute constant
$c_N>0$ with $\RR(F\mid k,l)\ge c_N\RR_N(F\mid k,l)\ge c_N2^{-k-2}$. With
$d:=\max(4/c_N,\,1/c)$ --- the second bound covering the case of finite $S$
settled at the beginning of the section --- we get
$\RR(F\mid k,l)\ge2^{-k}/d$, which is alternative~(ii).
\end{proof}


\section{Transfer from finite sets to all sets}\label{sec:transfer}

\begin{proof}[Proof of Theorem~\ref{th:transfer}]
Let $f$ be non-decreasing and computable and assume $\DD(F)<f(k)$ for all $k$ and
all finite $F$ with $\RR(F)\ge2^{-k}$. Let $C$ be a constant to be chosen at the
end of the proof, and given $k$ put
\[
  l\ :=\ f(k+2\log k+C)-1 .
\]
(We ignore rounding: $k+2\log k+C$ is to be read as
$k+\lceil2\log k\rceil+C$.) Note that $l$ is computed from $k$ by an algorithm
$g$, because $f$ is computable and $C$ is a constant; let $m$ be the length of a
program computing $g$.

Let $S$ be a set with $\RR(S)\ge2^{-k}$; we must show $\DD(S)<f(k+2\log k+C)$,
i.e.\ $\DD(S)\le l$. Suppose not, and apply Theorem~\ref{th:main} to $S$ with the
parameters $k$ and $l$. Alternative (i) fails, so there is a finite $F\subseteq
S$ with
\[
  \DD(F)>l \qquad\text{and}\qquad \RR(F\mid k,l)\ \ge\ 2^{-k}/d .
\]
We now remove the condition. Since $l=g(k)$, Lemma~\ref{lem:cond-basic}(c) gives
$\RR(F\mid k)\ge2^{-m-O(1)}\RR(F\mid k,l)$, and then
Lemma~\ref{lem:cond-basic}(b) gives
\[
  \RR(F)\ \ge\ c\,k^{-2}\RR(F\mid k)\ \ge\ c\,2^{-m-O(1)}\,k^{-2}\,2^{-k}/d
  \ \ge\ 2^{-(k+2\log k+c_1)} ,
\]
where $c_1=m+O(1)$; it depends only on $f$ and $C$ (through the algorithm $g$),
but not on $k$, $l$, $S$ or $F$. By the hypothesis of the theorem applied to the
finite set $F$ with the parameter $k+2\log k+c_1$,
\[
  \DD(F)\ <\ f(k+2\log k+c_1)\ \le\ f(k+2\log k+C)\ =\ l+1 ,
\]
provided $C\ge c_1$, where we used that $f$ is non-decreasing. This contradicts
$\DD(F)>l$.

It remains to check that $C$ can indeed be chosen with $C\ge c_1$. We have
$c_1=m+O(1)$, and $m$ --- the length of a program computing
$k\mapsto f(k+2\log k+C)-1$ --- is $a_f+O(\log C)$, where $a_f$ depends on $f$
alone and the term $O(\log C)$ accounts for writing the constant $C$ inside the
program. Hence $c_1\le a'_f+b\log C$ for an absolute constant $b$, and any
sufficiently large $C$ satisfies $C\ge a'_f+b\log C$.
\end{proof}

\begin{proof}[Proof of Corollary~\ref{cor:alpha}]
Let $\beta,\gamma$ be constants with
$\DD(F)\le\alpha\log\frac1{\RR(F)}+\beta\log\log\frac1{\RR(F)}+\gamma$ for all
finite $F$, and put
\[
  f(k)\ :=\ \lceil\alpha k\rceil+\lceil\beta\log k\rceil+\gamma+1 .
\]
This is a non-decreasing computable function (a fixed function, even though we
may not know $\alpha$, $\beta$ and $\gamma$), and for a finite $F$ with $\RR(F)\ge2^{-k}$
we have $\log\frac1{\RR(F)}\le k$, hence $\DD(F)<f(k)$. By
Theorem~\ref{th:transfer}, every $S$ with $\RR(S)\ge2^{-k}$ satisfies
\[
  \DD(S)\ <\ f(k+2\log k+c)\ =\ \alpha k+O(\log k) .
\]
Applying this with $k:=\lceil\log(1/\RR(S))\rceil$ gives
$\DD(S)\le\alpha\log\frac1{\RR(S)}+O\bigl(\log\log\frac1{\RR(S)}\bigr)$.
\end{proof}

\begin{proof}[Proof of Corollary~\ref{cor:two}]
By~\cite{Ver07}, $\DD(F)\le2\log\frac1{\RR(F)}+O(\log\log\frac1{\RR(F)})$ for
every finite $F$ (the bound there is stated for the prefix version of $\DD$,
which is an upper bound for ours by Remark~\ref{rem:notation}). Apply Corollary~\ref{cor:alpha} with $\alpha=2$.
\end{proof}

\begin{proof}[Proof of Corollary~\ref{cor:hardfinite}]
Let $\RR(S)>0$; then $\DD(S)<\infty$ (this is the theorem of~\cite{dLMSS56}, and
also follows from Corollary~\ref{cor:two}). If $\DD(S)=0$ --- which happens for
at most one set --- we may take $F=\varnothing$: then $\DD(F)\ge0=\DD(S)$ and
$-\log\RR(F)=O(1)$, since $\RR(\varnothing)>0$. So assume $\DD(S)\ge1$. Put
$l:=\DD(S)-1$ and
$k:=\lceil\log(1/\RR(S))\rceil$, so that $2^{-k}\le \RR(S)<2^{-k+1}$. Apply
Theorem~\ref{th:main} with these $k,l$. Alternative (i) would say
$\DD(S)\le\DD(S)-1$, so alternative (ii) holds and yields a finite $F\subseteq S$
with $\DD(F)>\DD(S)-1$, i.e.\ $\DD(F)\ge\DD(S)$, and
$\RR(F\mid k,l)\ge2^{-k}/d_0>\RR(S)/(2d_0)$, where $d_0$ is the constant of
Theorem~\ref{th:main}.

It remains to remove the condition. By Corollary~\ref{cor:two} we have
$l<\DD(S)\le 2k+O(\log k)$, so the pair $(k,l)$ has a self-delimiting encoding of
length $2\log k+2\log l+O(1)=O(\log k)$, and Lemma~\ref{lem:cond-basic}(b) gives
\[
  \RR(F)\ \ge\ c\,2^{-O(\log k)}\RR(F\mid k,l)\ \ge\ \RR(S)/k^{O(1)} .
\]
Since $k=\lceil\log(1/\RR(S))\rceil$, this is the second inequality of the
corollary.
\end{proof}

\section{Concluding remarks}

\subsection{What the new proof avoids}

Solovay's reduction and its subsequent refinements approximate $\RR(A)$ for an infinite $A$ by truncated quantities
$\RR^{t}_n(A)$ --- the probability that after $t$ steps the machine has
enumerated $A\cap[n]$ and nothing outside $A$ --- and must include in the
deterministic description a sequence of times $(t_n)$ at which these
approximations are reliable. That sequence costs $-\log\RR(A)$ extra bits, which
is exactly the loss in Theorem~\ref{th:solovay-red}. In the proof above no such
sequence is needed: the freezing mechanism makes the construction wait for the
probability mass instead of being told when to trust it, and the price is
reduced to the two numbers $k$ and $l$, supplied as a condition. Removing that
condition costs only $O(\log k)$ bits by Lemma~\ref{lem:cond-basic}: in the proof
of Theorem~\ref{th:transfer} because there $l$ is computable from $k$, and in
Corollary~\ref{cor:hardfinite} because there $l=O(k)$. It is also worth noting
that the argument never needs to know $\RR(A)$, nor to enumerate the sets $A$
under consideration.

\subsection{The Cats versus Ants game}\label{sec:game}

For completeness we recall the game of~\cite{Ver08} and the exact form of the
reduction, in the notation of the present paper.

\begin{definition}[Ants versus cats]\label{def:game}
The game is specified by a rational $\varepsilon\in(0,1)$ with $1/\varepsilon$
integer and by a number $k$ of cats. The nodes are the finite subsets of $\N$,
with an arc from $A$ to $B$ whenever $A\subsetneq B$. The ants form an infinitely
divisible unit mass; the $k$ cats are indivisible. At the start all the ants and
all the cats are at the node $\varnothing$. Bob moves the ants, Alice moves the
cats, and they move alternately, Bob first. In his turn Bob may perform finitely
many actions of the form ``move any portion of the ants from a node $A$ to a node
$B$ reachable from $A$''; in her turn Alice may perform finitely many actions of
the form ``move a cat from a node $A$ to a node $B$ reachable from $A$''. The
game lasts forever, and \emph{Alice wins} if after each of her moves every node
carrying at least $\varepsilon$ of the ants contains a cat.

Let $K(\varepsilon)$ be the least $k$ for which Alice has a computable winning
strategy in the $(\varepsilon,k)$-game, the strategy receiving $\varepsilon$ as
input.
\end{definition}

Note that a position of the game is a finite object: only finitely many nodes
carry ants or cats at any moment, and the ant mass on each of them is a rational
number, since each move redistributes finitely many portions. The infinite binary
strings of Section~\ref{sec:proof}, which index the ants there, play no role
here.

Bob wins whenever $k<1/\varepsilon$: on his first move he puts a portion
$\varepsilon$ of the ants in each of the nodes $\{1\},\dots,\{1/\varepsilon\}$,
and Alice has too few cats. Hence $K(\varepsilon)\ge1/\varepsilon$; in fact
$1/\varepsilon$ cats do not suffice either, as Bob wins the
$(1/4,4)$-game~\cite{Ver08}. On the other hand Alice wins with
$O(1/\varepsilon^{2})$ cats, by a computable strategy adapted from Martin's
strategy~\cite{Solovay77,Ver08}; so
\[
  1/\varepsilon\ <\ K(\varepsilon)\ =\ O\bigl(1/\varepsilon^{2}\bigr).
\]
What happens for $k\approx(1/\varepsilon)^{\alpha}$ with $1<\alpha<2$ is unknown.
The game is tied to our problem as follows.

\begin{theorem}[\cite{Ver08}]\label{th:game}
Let $\alpha\ge1$ be rational.
\begin{enumerate}
\item[(1)] If for some constant $c$ Alice has a computable strategy that,
  given $\varepsilon$, wins the $(\varepsilon,k)$-game with some $k\le c\,(1/\varepsilon)^{\alpha}$ for all $\varepsilon$, then
  $\DD(F)\le\alpha\log\frac1{\RR(F)}+O\bigl(\log\log\frac1{\RR(F)}\bigr)$
  for every finite $F\subseteq\N$.
\item[(2)] Conversely, if for some $\delta>0$ and arbitrarily small
  $\varepsilon$ Bob has a computable winning strategy in the
  $(\varepsilon,\delta(1/\varepsilon)^{\alpha})$-game, then
  $\DD(F)\ge\alpha\log\frac1{\RR(F)}-O(\log \DD(F))$ for infinitely many finite
  $F\subseteq\N$.
\end{enumerate}
\end{theorem}

Thus the game bounds the optimal constant for finite sets from both sides: by
(1) every $\alpha$ for which Alice wins computably is an upper bound, and by~(2) every $\alpha$ for which
Bob wins computably is a lower bound.  Note that if Bob has a winning strategy, then he also has a computable winning strategy. This follows from the fact that his strategy wins in a finite number of moves (if Alice can prolong the game forever, then this is not a valid winning strategy for Bob). This argument requires that in every position Alice have only finitely many essentially different moves, and this is indeed the case. Let $U$ be the union of all nodes carrying ants after Bob's move. A cat at a node $B\not\subseteq U$ covers no node carrying ants, and every node reachable from $B$ is also reachable from $B\cap U$. Hence moving a cat to $B\cap U$ instead of $B$ never hurts Alice, and we may assume that she moves cats only to subsets of~$U$ (then all cats always sit at subsets of~$U$, so such moves are legal). Thus Alice has only finitely many moves in each position, and by K\"onig's lemma every winning strategy for Bob wins within a bounded number of moves, that is, it is a finite object. Note that Bob has infinitely many moves, but they are finite objects (finitely many portions with rational masses) and can be effectively enumerated; K\"onig's lemma is applied to the tree of plays consistent with a fixed strategy of Bob, which branches only at Alice's moves. By bruteforcing all strategies that win in a finite time, Bob can find the winning one. Therefore the requirement of computability in Theorem~\ref{th:game}(2) can be dropped: it suffices that for some $\delta>0$ and arbitrarily small $\varepsilon$ Bob has a winning strategy in the $(\varepsilon,\delta(1/\varepsilon)^{\alpha})$-game. However, we do not know if every winning strategy of Alice can be made computable. Since the game is determined (the winning condition for Alice is closed), this is the only obstacle: if Alice, whenever she can win with $c\,(1/\varepsilon)^{\alpha}$ cats for all $\varepsilon$, could also do this by a single computable strategy receiving $\varepsilon$ as input, the two bounds would meet. As it stands, we cannot conclude that they meet. 

Our transfer theorem shows that the game governs
\emph{all} sets, not only finite ones:

\begin{corollary}\label{cor:game-all}
If Alice has a computable strategy as in Theorem~\ref{th:game}(1) for a rational
$\alpha\ge1$, then
$\DD(S)\le\alpha\log\frac1{\RR(S)}+O\bigl(\log\log\frac1{\RR(S)}\bigr)$ for
every $S\subseteq\N$, finite or infinite.
\end{corollary}

\begin{proof}
Combine Theorem~\ref{th:game}(1) with Corollary~\ref{cor:alpha}.
\end{proof}

For $\alpha=2$ this is a second route to Corollary~\ref{cor:two}: Alice's
strategy with $O(1/\varepsilon^{2})$ cats gives the bound for finite sets, and
Theorem~\ref{th:transfer} carries it over to all sets. In particular, any
improvement of Alice's strategy would immediately improve the bound for arbitrary
sets.

\subsection{The remaining question}

By Corollary~\ref{cor:alpha} the optimal constant $\alpha$ is the same for finite
and for arbitrary sets, and by Theorem~\ref{th:game} it is bounded above by every exponent for which Alice has computable winning strategies and below by every exponent for which Bob has
 winning strategies. We know $1\le\alpha\le2$. The upper bound $2$ comes from
Martin's game, where the analogous quantity satisfies
$N(\varepsilon)=\Theta((1/\varepsilon)^{2})$~\cite{Solovay77,Ageev02}, so no
improvement can come from that game; for Cats versus Ants only the trivial lower
bound is known. Determining $\alpha$ --- in particular deciding whether
$\alpha=1$, i.e.\ whether randomness helps only by an additive logarithmic term,
as it does in the halting case --- remains open.

\section{AI disclosure}

In preparing this paper the authors used a large language model (Claude,
Anthropic) to speed up the writing: it was used to draft and polish the
English prose, to help with \LaTeX{} formatting, and to check the text for
consistency of notation and for typos. All results of the paper --- the
definitions, theorems and their proofs --- were obtained by the authors
themselves without any use of generative models. All text produced with the
assistance of the model was checked by the authors, who take full
responsibility for its correctness.


\begin{thebibliography}{99}

\bibitem{Chaitin75} G.~J. Chaitin. A theory of program size formally identical
  to information theory. \emph{Journal of the ACM} 22(3):329--340, 1975.

\bibitem{Ageev02} M.~Ageev. Martin's game: a lower bound for the number of sets.
  \emph{Theoretical Computer Science} 289(1):871--876, 2002.

\bibitem{dLMSS56} K.~de Leeuw, E.~F.~Moore, C.~E.~Shannon, N.~Shapiro.
  Computability by probabilistic machines. In \emph{Automata Studies},
  pages 183--212, Princeton University Press, 1956.

\bibitem{Levin74} L.~A.~Levin. Laws of information conservation (non-growth) and
  aspects of the foundation of probability theory. \emph{Problems of Information
  Transmission} 10(3):206--210, 1974.

\bibitem{LV08} M.~Li, P.~Vit\'anyi. \emph{An Introduction to Kolmogorov
  Complexity and Its Applications.} Springer, 3rd edition, 2008.

\bibitem{SUV} A.~Shen, V.~A.~Uspensky, N.~Vereshchagin. \emph{Kolmogorov
  Complexity and Algorithmic Randomness.} Mathematical Surveys and Monographs,
  vol.~220, American Mathematical Society, 2017.

\bibitem{Solovay77} R.~M.~Solovay. On random r.e.\ sets. In \emph{Non-Classical
  Logics, Model Theory and Computability}, volume 89 of \emph{Studies in Logic
  and the Foundations of Mathematics}, pages 283--307, North-Holland, 1977.

\bibitem{Ver07} N.~K.~Vereshchagin. Kolmogorov complexity of enumerating finite
  sets. \emph{Information Processing Letters} 103(1):34--39, 2007.

\bibitem{Ver08} N.~K.~Vereshchagin. Kolmogorov complexity and games.
  \emph{Bulletin of the EATCS} (94), 2008.

\end{thebibliography}
\end{document}